\documentclass[aps,prl,reprint,groupedaddress]{revtex4-1}

\usepackage{graphicx,amsmath,color}

\newcommand{\beq}{\begin{equation}}
\newcommand{\eeq}{\end{equation}}

\begin{document}

\title{Thermionic Emission and Negative $dI/dV$ in Photoactive Graphene Heterostructures}

\author{J. F. Rodriguez-Nieva$^{1}$, M. S. Dresselhaus$^{1,2}$, L. S. Levitov$^{1}$}
\affiliation{${}^1$Department of Physics, Massachusetts Institute of Technology, Cambridge, MA 02139, USA}
\affiliation{${}^2$Department of Electrical Engineering and Computer Science, Massachusetts Institute of Technology, Cambridge, MA 02139, USA}
\date{\today}
\begin{abstract}
Transport in photoactive graphene heterostructures, originating from the dynamics of photogenerated hot carriers, is governed by the processes of thermionic emission, electron-lattice thermal imbalance and cooling. These processes give rise to interesting photoresponse effects, in particular negative differential resistance (NDR) arising in the hot-carrier regime. The NDR effect stems from a strong dependence of electron-lattice cooling on the carrier density, which results in the carrier temperature dropping precipitously upon increasing bias. The ON-OFF switching between the NDR regime and the conventional cold emission regime, as well as the gate-controlled closed-circuit current that is present at zero bias voltage, can serve as signatures of hot-carrier dominated transport.

\end{abstract}




\maketitle

Graphene, because of its unique characteristics, is of keen interest for optoelectronics research in areas such as photodetection, solar cells and light-emitting devices \cite{optoelectronics,optoelectronics2,photodetector,lightabs}. Recently it was emphasized that graphene features an unusual kind of photoresponse mediated by photogenerated hot carriers. The hot-carrier regime originates from the quenching of electron-lattice cooling when the system is close to charge-neutrality \cite{coolingmcd,coolingdassarma}. The hot carriers are exceptionally long-lived in graphene and can proliferate across the entire system \cite{pteffect,pteffect2,photocurrentscience}. This behavior, which sets graphene apart from other photoactive materials, leads to a dramatic enhancement in photo-response. 

\begin{figure}[t]
\centering \includegraphics[scale=1.0]{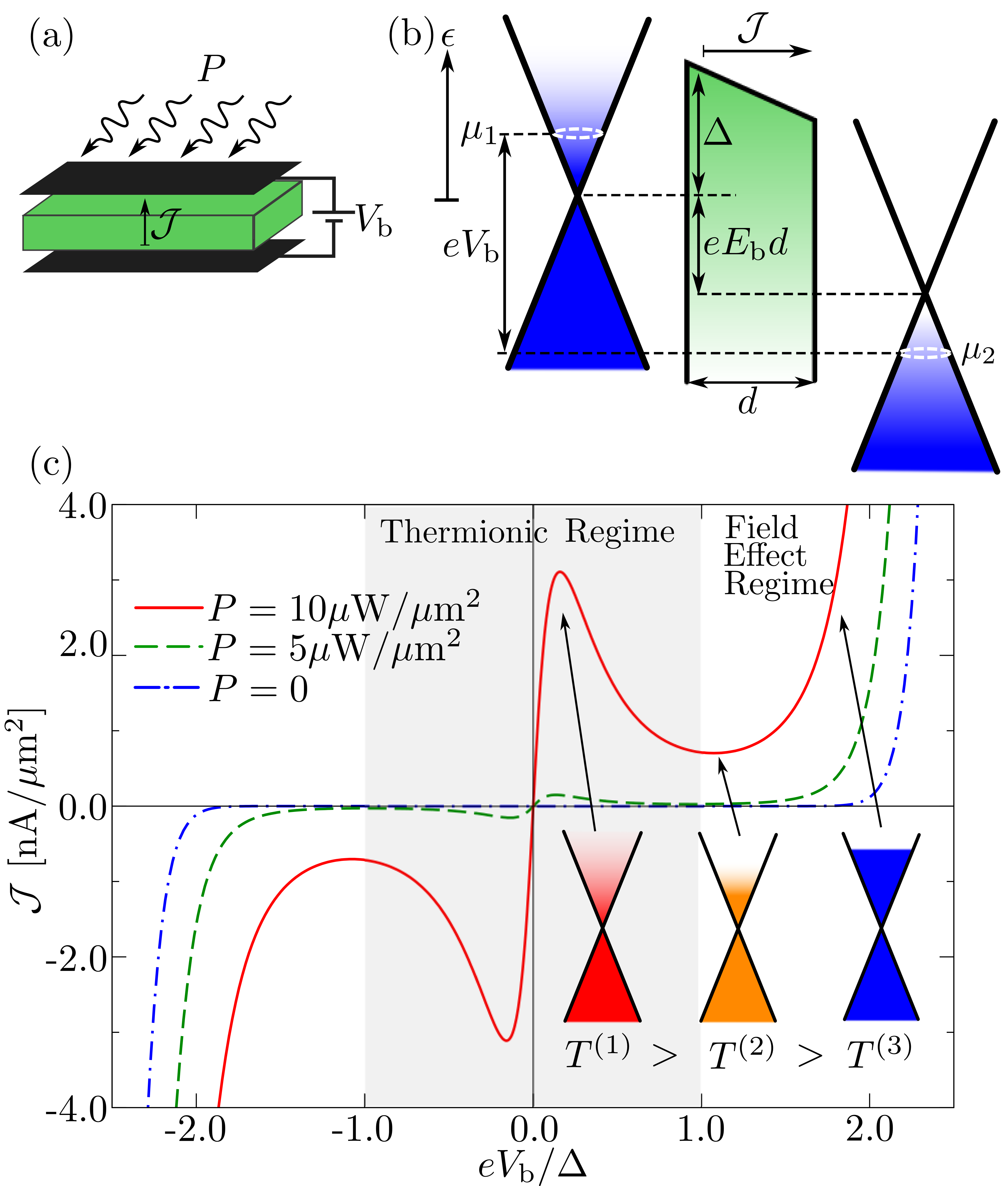}
\caption{The Negative Differential Resistance (NDR) effect in a photoactive heterostructure operated in the thermionic emission regime. Shown here are (a) device schematics,  (b) electronic band structure with the quantities discussed in the text marked, and (c) the $I-V$ dependence under optical pumping obtained from Eqs.\,(\ref{eq:current})--(\ref{eq:qc}). The bias voltage $V_{\rm b}$ controls the electron cooling through electrostatic doping of graphene layers. An enhancement in the cooling power upon increased $V_{\rm b}$ triggers carrier temperature dropping [marked $T^{(1,2,3)}$ in (c)]; see also simulation results in Fig.\ref{fig:disorder}(a). The resulting suppression of thermionic current leads to {\it negative} $dI/dV$ (in the grey region). For larger bias values, transport is dominated by field emission, yielding {\it positive} $dI/dV$ outside the grey region. Shown here are results for both  graphene layers undoped at $V_{\rm b}=0$. Results for nonzero doping are presented in Figs.\ref{fig:jsc},\ref{fig:currentmap}. 
}
\label{fig:current}
\end{figure}

The 2D character of electronic states, which are fully exposed in materials such as graphene, can enable new device architectures. One system of high current interest is stacked graphene-dielectric-graphene structures [see Fig.\ref{fig:current}(a)]. Fabricated with atomic precision, such systems can behave as field-effect transistors \cite{tunnelingscience,tunnelingbn}, resonant tunnel diodes \cite{ndrtransistor,ndrtransistor2}, photodetectors \cite{photodetector1,photodetector2}; they also provide a platform to explore the Coulomb Drag effect \cite{drag2} and the metal-insulator transition \cite{gmit}. In the hot-carrier regime, proliferation of photoexcited electron-hole pairs can result in an enhanced thermionic emission of hot carriers over the barrier. Vertical carrier extraction in such structures is facilitated by short interlayer transport lengths in the nanometer range, ultra-fast response times, and large active areas. Variability in properties of different 2D barrier materials (hBN, MoS${}_2$, WSe${}_2$, etc.) allows to tailor the photo-response characteristics to the different optoelectronic applications. 

Here we predict that interlayer transport in graphene heterostructures operating in the hot-carrier regime leads to an unusual type of photoresponse, namely, a negative differential resistance (NDR). The mechanism of this NDR response relies on the interplay of two effects. First, the phase space available for phonon scattering rapidly increases with doping, enhancing the electron-lattice cooling and thereby altering the number of hot carriers in the system. Second, the large capacitance of the atomically thin device renders the carrier density in graphene layers sensitive to the interlayer potential difference. These two effects combined together result in a reduction of the electronic temperature and a suppression of thermionic current upon an increase of the bias potential $V_{\rm b}$. The NDR effect arises when this suppression overwhelms the increase in the field-effect transport under bias. Such an NDR mechanism manifests itself as an enhanced photo-current peaked at a bias potential well below the onset of the conventional field-emission regime, $eV_{\rm b}\ll\Delta$, where $\Delta$ is the barrier height, as shown in Fig.\ref{fig:current}(c).

We note that the photoactive NDR architectures are a class of their own, and are well suited for optoelectronic applications. In particular, the fast response and  \textit{in-situ} tunability of graphene devices makes them ideal as photo-active switches or light-detectors with high gain. The NDR effect in photo-active devices, analyzed below, is distinct from the one in traditional NDR devices, such as Gunn diodes \cite{gunn} or resonant tunneling diodes \cite{resonanttunneling,resonanttunnelingexp} which rely on non-linearities under the application of large electric fields in the absence of light.

Turning to the technical discussion, we note that, while in general both electrons and holes can contribute to thermionic transport, in practice transport is often dominated by a single carrier type. In the case of hBN, the barrier heights are $\Delta_{\rm el}\sim 3.5\,{\rm eV}$ for electron transport and $\Delta_{\rm h}\sim 1.3\,{\rm eV}$ for hole transport \cite{barrierheight}. We can therefore treat the interlayer transport in an hBN-based system as dominated by a single carrier type (holes). 

Below we focus on the behavior in wide-barrier structures, where thermionic emission of thermally activated carriers dominates over direct tunneling. This is the case for hBN thicknesses exceeding 4-5 monolayers ($d\sim 1\,{\rm nm}$) \cite{tunnelingbn} at sufficiently high temperatures. Thermionic currents are described by a particularly simple model when both graphene layers are at neutrality at $V_{\rm b}=0$\cite{thermionic3D}:
\beq
{\cal J}(V_{\rm b},T)= (g_0/e\beta) e^{-\beta\Delta}\sinh (\beta eV_{\rm b}/2)
,
\label{eq:thermalcurrent}
\eeq
where ${\cal J}$ is the current density  per unit area, $\beta^{-1}=k_{\rm B}T$ and the $g_0$ value is estimated in Eq.(\ref{eq:G0}). This expression follows from a general microscopic model at not too high bias $V_{\rm b}$, such that the effect of anti-symmetric doping induced by $V_{\rm b}\ne 0$ is stronger than the corresponding change in the barrier skewness [see derivation and discussion in the paragraph before Eq.(\ref{eq:condition})]. For larger bias values, field corrections to the barrier potential become important and must be accounted for; this is done in a microscopic model developed below. We also note that the electronic distribution is typically non-exponential when relaxation is slow. Slow relaxation would make the distribution tails more pronounced, ultimately enhancing the thermionic effects.

The steep dependence of ${\cal J}$ on $T$ and $V_{\rm b}$ in Eq.(\ref{eq:thermalcurrent}) leads to NDR by the following mechanism. For electrons in thermal equilibrium with the lattice, Eq.\,(\ref{eq:thermalcurrent}) predicts a monotonic $I-V$ dependence. A very different behavior, which is key for NDR, arises under pumping. As pictured schematically in Fig.\ref{fig:current}(c) for three values $T^{(1)}>T^{(2)}>T^{(3)}$, in the hot-carrier regime the electron temperature $T$ becomes highly sensitive to $V_{\rm b}$. The temperature-bias coupling arises because of the large interlayer capacitance producing bias-dependent doping in the graphene layers. An increase in carrier density leads to a faster electron-lattice cooling, which reduces thermal imbalance $\Delta T=T-T_{0}$, with $T_0$ the lattice temperature. The dependence in Eq.(\ref{eq:thermalcurrent}) then predicts suppression of thermionic emission.  If strong enough, this suppression can lead to negative $dI/dV$. The NDR effect takes place in the grey region marked in Fig.\ref{fig:current}(c). 

\begin{figure}
\centering \includegraphics[scale=1.0]{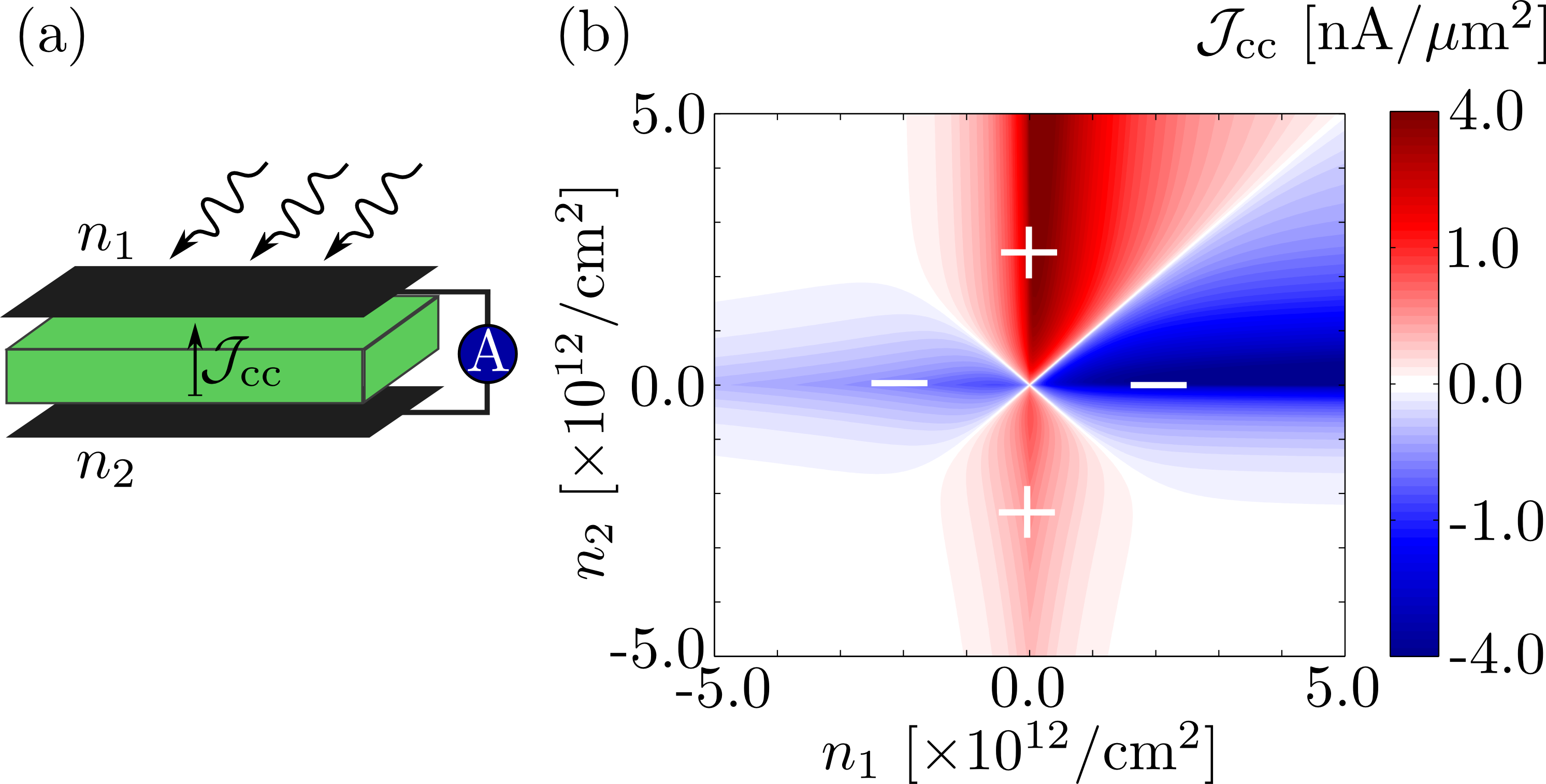}
\caption{
The closed-circuit current 
${\cal J}_{\rm cc}$ induced by optical pumping ($P=10\,\mu{\rm W}/\mu{\rm m^2}$) in the absence of voltage bias, $V_{\rm b}=0$. In the transport regime dominated by hot carriers, the value and polarity of  ${\cal J}_{\rm cc}$ are sensitive to the carrier densities $n_1$, $n_2$ in the graphene layers. Shown are experimental schematics (a), and the dependence ${\cal J}_{\rm cc}$ {\it vs.} $n_1$ and $n_2$ (b). The four-fold pattern with multiple changes of the current polarity arises due to the strong cooling power dependence on carrier concentration (a non-linear color scale is used to amplify the features of interest). 
}
\label{fig:jsc}
\end{figure}

The sensitivity of the electron-lattice cooling to carrier concentration provides a smoking gun for the regime dominated by hot carriers, helping distinguish it  from the conventional resonant tunneling NDR mechanisms, such as those discussed in Refs.\,\cite{ndrtransistor,ndrtransistor2}. In that regard, we mention that for the more general case of unequal carrier densities ($n_1\ne n_2$ at $V_{\rm b}=0$), our analysis predicts that the interlayer thermionic transport persists even at zero bias. 
As illustrated in Fig.\ref{fig:jsc}, this produces a closed-circuit current at $V_{\rm b}=0$ with a characteristic density dependence: a four-fold pattern with multiple polarity changes. Such a pattern provides a characteristic signature of hot-carrier dominated transport. 

\begin{figure}
\centering \includegraphics[scale=1.0]{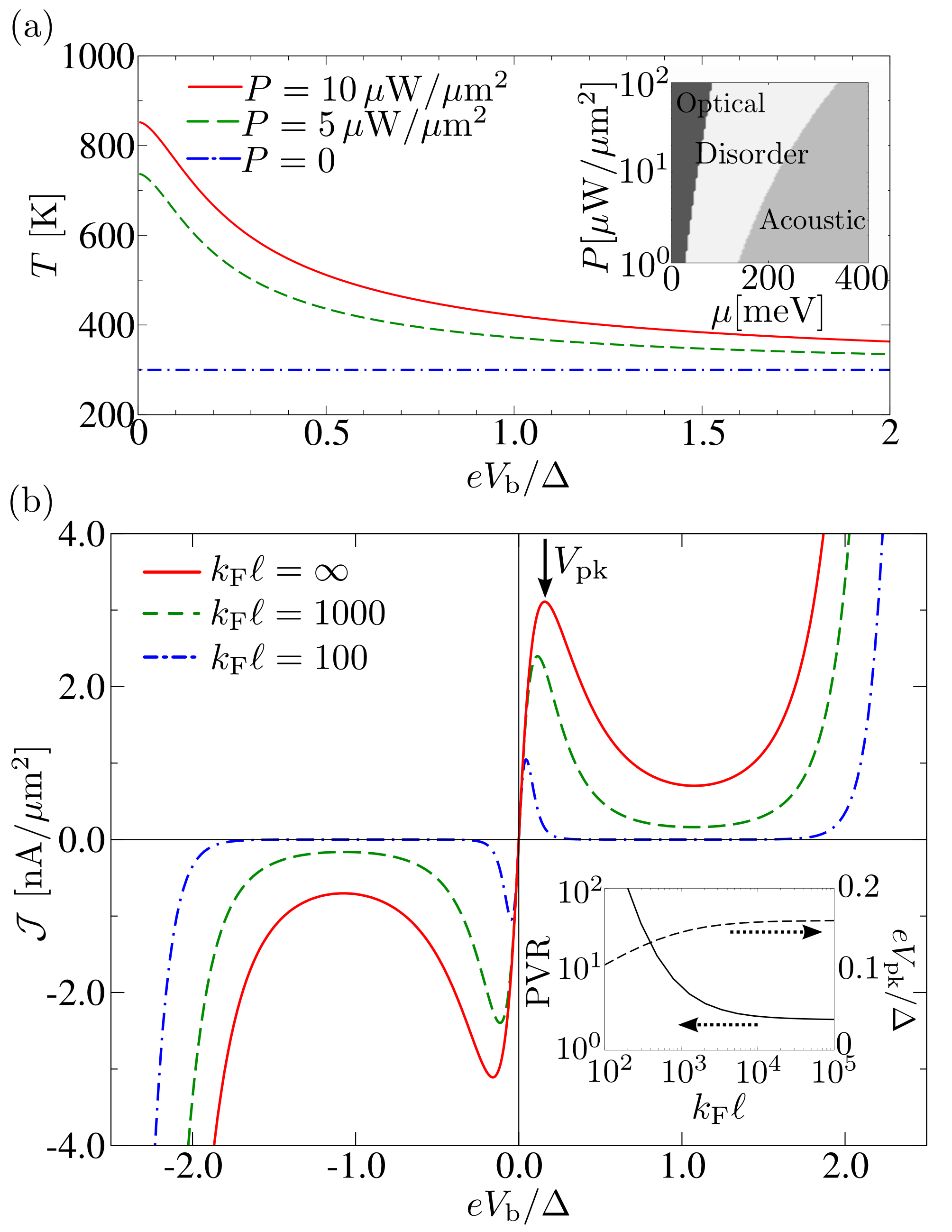}
\caption{(a) Electronic temperature $T$ {\it vs.} bias $V_{\rm b}$ for different pump power values $P$. 
The inset shows the dominant cooling pathways for different chemical potential and pump power values [for $T_{0}=0\,{\rm K}$, $k_{\rm F} \ell = 100$]. (b) Current ${\cal J}$ {\it vs.} $V_{\rm b}$ for different $k_{\rm F} \ell$ values, and $P=10\,\mu{\rm W}/\mu{\rm m^2}$. The bias potential at which current peaks ($V_{\rm pk}$) and the peak-to-valley ratio (PVR) are sensitive to the amounts of disorder (see inset). Results shown in this figure (except the inset of panel a) are obtained for the system undoped at $V_{\rm b}=0$, as in Fig.\ref{fig:current}.} 
\label{fig:disorder}
\end{figure}

The strength of the hot-carrier effects, reflected in the response of $T$ to $V_{\rm b}$ [see Fig.\ref{fig:disorder}(a)], can be characterized by the dimensionless quantity
\beq
\alpha=-\frac{\Delta}{T} \frac{dT}{d(eV_{\rm b})}.
\label{eq:alpha}
\eeq
The value $\alpha$
depends mainly on the power $P$ pumped into the electronic system and on the thickness $d$ of the barrier via the capacitance effect. 
As we will see, $\alpha$ governs the N-shaped $I-V$ dependence: the condition for NDR can be stated as $\alpha > 1/2$. We will argue that values as large as $\alpha\sim 25$ can be reached under realistic conditions.

We now proceed to introduce our transport model. For thermionic transport over the barrier, as well as for tunneling through it, we adopt a quasi-elastic but momentum non-conserving approximation. Indeed, a number of momentum scattering mechanisms at the interface are possible, such as scattering by defects, intrinsic phonons, substrate phonons, {\it etc}. Typical energy exchange in these processes is small on the barrier height scale $\Delta$. With this in mind, we use the (quasi-elastic) WKB model for the interlayer transition matrix element: 
\beq
t(\epsilon)=\Gamma e^{-\frac1{\hbar}\int_{0}^{x_{*}} p(x)dx}
,\quad
\frac{p^2(x)}{2m}=\Delta - eE_{\rm b}x-\epsilon, 
\label{eq:wkb}
\eeq
for $\epsilon<\Delta$, and $t(\epsilon)=\Gamma$ for $\epsilon>\Delta$, where $\Gamma$ is an energy-independent prefactor which depends on the barrier material properties. In Eq.\,(\ref{eq:wkb}), $E_{\rm b}$ is the electric field within the barrier due to interlayer bias, see Eq.(\ref{eq:qc}), $m$ is the electron effective mass in the dielectric, and $x_{*}$ is the classical turning point for the skewed barrier potential, $x_{*}=\min[d,(\Delta-\epsilon)/eE_{\rm b}]$, see Fig. \ref{fig:current}(b). Here, for the sake of simplicity, we ignore the effect of attraction to image charges, described by a $-1/|x|-1/|x-d|$ potential. For a large barrier width, this gives rise to the barrier height Schottky dependence on the square root of $E_{\rm b}$. The effect is less dramatic for the not-so-large barrier widths analyzed below.

Assuming an elastic but momentum non-conserving interlayer transport, the vertical current ${\cal J}(V_{\rm b})$ can be expressed through barrier transmission and carrier distribution\cite{wolf}:
\beq
{\cal J}(V_{\rm b})= e\sum_{\epsilon,ij}\frac{2\pi}{\hbar}|t_{ij}(\epsilon)|^2 D_{1}(\epsilon)D_{2}(\tilde\epsilon) [f_1(\epsilon)-f_2(\tilde\epsilon)]
,
\label{eq:current}
\eeq
where, due to the built-in field between layers,  the energy for the quantities in layer 2 is offset by $\tilde\epsilon=\epsilon+eE_{\rm b}d$ [see Fig.\ref{fig:current}(b)]. 
Here ${\cal J}$ is the current per unit area, $f_1(\epsilon)=[e^{\beta_1(\epsilon-\mu_1)}+1]^{-1}$, $f_2(\epsilon)=[e^{\beta_2(\epsilon-\mu_2)}+1]^{-1}$
are the Fermi distribution functions,
the sum 
denotes integration over $\epsilon$ and summation over spins and valleys. We use $t_{ij}(\epsilon)=\delta_{ij}t(\epsilon)$ defined in Eq.(\ref{eq:wkb}), and  the density of states per spin/valley $D_{1,2}(\epsilon)=|\epsilon|/2\pi(\hbar v )^2$ with $v\approx 10^6\,\rm{m/s}$ the carrier velocity. The temperatures established under pumping in unequally doped layers are generally distinct, $\beta_1\ne\beta_2$, reflecting the cooling rates density dependence. Also, importantly, the electrostatic potential between layers, $E_{\rm b}d$, is distinct from the bias voltage $V_{\rm b}$. This is so because the quantum capacitance effects, prominent at small carrier densities,\cite{qcapacitance}  make the quantites $\mu_{1}$ and $\mu_{2}$ bias-dependent. These effects will be investigated below. 

A feature of our system which is key for NDR is the large mutual capacitance of graphene layers, which couples $V_{\rm b}$ with the carrier density and makes the hot-carrier properties of each layer tunable. This coupling acts as a knob producing big changes in the hot-carrier photoresponse through modest changes of carrier concentration on the order $\delta n \sim 10^{12}\,{\rm cm}^{-2}$. Such bias-induced doping changes are routine in graphene/hBN systems \cite{tunnelingscience}. 

The bias-induced changes in carrier densities of graphene layers, as well as the electric field $E_{\rm b}$ between the layers, can be described by a simple electrostatic model. We consider a dual-gated device with fixed charge densities $n_{\rm T}$, $n_{\rm B}$ in the top and bottom gates, respectively. The neutrality condition relates the charge densities in the different regions of the device as
\beq
n_{\rm T}+n_{1}+n_{2}+n_{\rm B}=0
,
\label{eq:neutrality}
\eeq
where $n_{1}$, $n_{2}$ are carrier densities on the graphene layers. The field $E_{\rm b}$ is related to these quantities through Gauss' law: 
\beq
\kappa  E_{\rm b}=2\pi e \,(n_{\rm T}+n_{1}-n_{2}-n_{\rm B}),
\label{eq:Eb}
\eeq
where $\kappa $ is the dielectric constant of the barrier. A bias voltage $V_{\rm b}$ applied between the graphene layers results in 
\beq
eV_{\rm b}=\mu_1-\mu_2+eE_{\rm b}d.
\label{eq:qc}
\eeq
With doping-dependent $\mu_{1,2}$, Eq.(\ref{eq:qc}) accounts for the quantum capacitance effects. Here we will use the $T=0$ expression $\mu_i={\rm sign}(n_i)\hbar v  \sqrt{\pi |n_i|}$, which provides a good model over most of the relevant carrier density range. 

The three unknown variables $n_1$, $n_2$ and $E_{\rm b}$ can now be found by solving the three equations (\ref{eq:neutrality})-(\ref{eq:qc}), once the external variables $n_{\rm T}$, $n_{\rm B}$ and $V_{\rm b}$ are fixed. Throughout this work we focus on the symmetric case when $n_{\rm T}=n_{\rm B}=-n_{0}$ with no interlayer bias  applied (which corresponds to both graphene layers at neutrality when the gates are uncharged). In this case, Eqs.\,(\ref{eq:neutrality}),(\ref{eq:Eb}) can be restated as $n_1=n_0+\delta n$, $n_2=n_0-\delta n$, and $\kappa  E_{\rm b}=4\pi e\delta n$. Then, plugging these values into Eq.\,(\ref{eq:qc}), the density imbalance $\delta n$ can be obtained. The built-in field $E_{\mathrm{b}}$ matters in two different ways: the electrostatic potential value $eE_{\mathrm{b}}d$ enters the WKB model, Eq. (\ref{eq:wkb}), as well as in the offset between $D_1(\epsilon)$ and $D_2(\epsilon)$ in Eq. (\ref{eq:current}). 

\begin{figure}
\centering \includegraphics[scale=1.0]{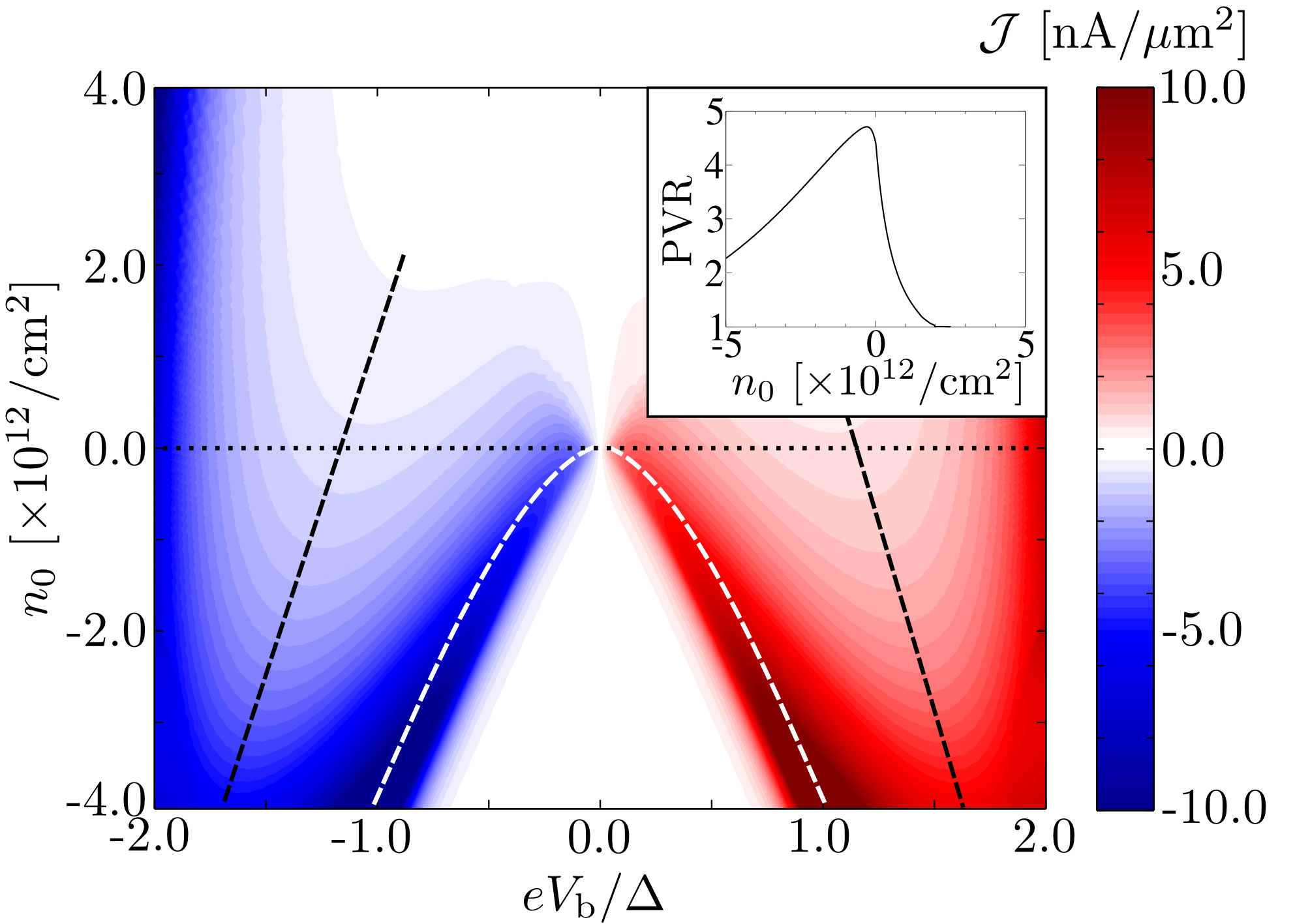}
\caption{Map of the current 
${\cal J}$ {\it vs.} $eV_{\rm b}/\Delta$ and zero-bias doping $n_{0}$ (equal for both layers), for $P=10\,\mu{\rm W}/\mu{\rm m^2}$. 
The current peak positions track the $V_{\rm b}$ values where the emitter is at neutrality (white dashed line). Local minima of
${\cal J}$ {\it vs.} $V_{\rm b}$ are marked with black dashed lines. 
The $n_0=0$ slice,  plotted in Fig.\ref{fig:current}(c), is marked with a dotted line.  The peak-to-valley ratio (PVR) {\it vs.} $n_0$ is shown in the inset.
}
\label{fig:currentmap}
\end{figure}

The $I-V$ dependence for the case $n_0=0$ in shown in Fig.\ref{fig:current}; for finite values $n_{\rm T}=n_{\rm B}=-n_0$ it is shown in Fig.\ref{fig:currentmap}. In our simulation, we first determine $\mu_{1,2}$ from Eqs.(\ref{eq:neutrality})-(\ref{eq:qc}) as a function of $V_{\rm b}$ and $n_0$. Using the $\mu_{1,2}$ values, the electronic temperatures $T_{1,2}$ are determined from energy balance considerations, see Eq.(\ref{eq:totalpower}) below. Finally, using the calculated values $\mu_{1,2}$, $T_{1,2}$, and $E_{\rm b}$, the current  is obtained from Eq.(\ref{eq:current}). We use hBN barrier parameters for numerical estimates, with a (hole) barrier height $\Delta \sim 1.3\,{\rm eV}$ \cite{barrierheight}, dielectric constant $\kappa  \sim 5 $, thickness $d=6\,{\rm nm}$ ($\sim$ 20 monolayers). Room temperature is assumed, $T_{0}=300\,{\rm K}$, unless stated otherwise. 

The prefactor value $\Gamma$ in Eq.(\ref{eq:wkb}) can be related to the measured conductance.
Physically, $\Gamma$ accounts for 
the processes in which tunneling couples to phonons or defects. The rates for these processes, which typically vary from interface to interface, can be estimated from transport measurements. 
Linearizing Eq.(\ref{eq:current}) in $eV_{\rm b}$ for $k_{\rm B} T \ll \Delta$, and accounting for the thermionic contribution due to $\epsilon\approx\Delta$, the zero-bias conductance per unit area is
\beq\label{eq:G0}
G = \frac{g_0}{2} e^{-\beta \Delta}
,\quad
g_0=4\pi N D_1(\Delta)D_2(\Delta)|\Gamma|^2 \frac{e^2}{\hbar},
\eeq
where we take $T_1=T_2$. Here $N=4$ is the spin/valley degeneracy, and $g_0$ is a prefactor in Eq. (\ref{eq:thermalcurrent}). 
The activation  $T$ dependence is consistent with that measured for dark current \cite{transistornature}. Comparison with values $G\sim 10^{-7}\,\Omega^{-1}\mu{\rm m^{-2}}$ measured at room temperature, and $\Delta\sim 0.4\,{\rm eV}$ \cite{transistornature}, yields values $g_0 \sim 1\,{\rm \Omega^{-1}\mu m^{-2}}$ and $\Gamma \sim 0.5\,{\rm eV\AA}$.

Next we discuss how hot-carrier effects result in a coupling between the electronic temperature and the chemical potential for each  graphene layer. In the continuous wave regime, the power $P$ pumped into the electronic system is distributed among the electron and lattice degrees of freedom. For simplicity, we will use a two-temperature model, describing electrons by a temperature distinct from the lattice temperature, $T>T_0$, valid when the carrier-carrier scattering rate is faster than the electron-lattice relaxation rate. Assuming spatially uniform in-plane temperatures and chemical potentials, the total cooling power $P$ obeys the energy balance condition 
\beq
P=P_{\rm ac}(\mu_{i},T_{i})+P_{\rm opt}(\mu_{i},T_{i})+P_{\rm dis}(\mu_{i},T_{i})
,
\label{eq:totalpower}
\eeq
written separately for each layer $i=1,2$. Here we ignored effects such as direct interlayer energy transfer as well as the heat drained through the contacts. Equation (\ref{eq:totalpower}) accounts for three cooling pathways intrinsic to graphene, mediated by acoustic and optical phonons ($P_{\rm ac}$, $P_{\rm opt}$) \cite{coolingmcd,coolingdassarma}, and the disorder-assisted acoustic phonon mechanism  ($P_{\rm dis}$) \cite{disordercooling}, see Supplement. The cooling rates in Eq.(\ref{eq:totalpower}) also depend on the lattice temperature $T_{0}$, however it suffices to treat $T_{0}$ as a fixed parameter, since the heat capacity of the lattice greatly exceeds that of the electron system. 

The intralayer Joule heating is small and thus need not be included in Eq.\,(\ref{eq:totalpower}). Indeed, typical vertical current values obtained in devices of active area $\sim 1\,{\rm \mu m^2}$ and under a bias $V_{\rm b}\sim 1\,{\rm V}$ 
do not exceed a few nA \cite{tunnelingscience,tunnelingbn,transistornature}. This yields Joule heating sources which are at least three orders of magnitude smaller than the powers pumped optically.   The interlayer energy transfer as well as the heat drained through contacts can be ignored in the energy balance in Eq.\,(\ref{eq:totalpower}) for similar reasons.

The cooling power is a strong function of doping for the acoustic-phonon contributions $P_{\rm ac}$ and $P_{\rm dis}$. This is so because acoustic phonon scattering is dominated by quasi-elastic scattering processes at the Fermi surface, and also because of the strong dependence of the electron-phonon coupling on the phonon energy. The resulting dependence on the chemical potential takes the form\cite{coolingmcd,disordercooling}
\beq
P_{\rm ac}\propto \mu^{4}(T-T_{0})
,\quad
P_{\rm dis}\propto \mu^2(T^{3}-T_{0}^{3})/k_{\rm F} \ell
\eeq
 in the degenerate limit $k_{\rm B}T\ll \mu$. The factor $1/k_{\rm F} \ell$, where $\ell$ is the disorder mean free path, describes the dependence of $P_{\rm dis}$ on disorder strength. In contrast, the contribution $P_{\rm opt}$ is essentially $\mu$-independent. Since the optical phonon energy in graphene is quite large, $\hbar \omega_0 \sim 0.2\,{\rm eV}$, the value $P_{\rm opt}$ is quite small, behaving as $P_{\rm opt}\propto \exp(-\hbar\omega_{0}/T)$ for $k_{\rm B}T,\mu \lesssim \hbar \omega_{0}$. The ratio $P_{\rm ac}/P_{\rm opt} $ is therefore small near charge neutrality but can be order-one for strongly doped graphene with typical doping $n \sim 10^{13}\,{\rm cm}^{-2}$ \cite{coolingmcd}. The cooling power strong dependence on $\mu$ can trigger the temperature dropping upon an increase in $\mu$, i.e. $dT/d\mu < 0$, see Fig.\ref{fig:disorder}(a).

The transition between the hot-carrier dominated regime, and the conventional field emission regime can be controlled by the power pumped into the electronic system. In our simulation we used the values for $P$ typical of laboratory lasers below saturation \cite{saturation} (a few mW per a ${\rm \mu m}$-wide spot). We take $P$ to represent the power {\it absorbed} in each graphene layer (2.3\% of incident power \cite{lightabs}). The $P$ values used are quoted in Figs.\ref{fig:current},\ref{fig:disorder} panels and in Figs.\ref{fig:jsc},\ref{fig:currentmap} captions. As shown in Fig.\ref{fig:current}(c), a strictly monotonic $I-V$ response obtained at $P=0$, transforms into an N-shaped NDR dependence upon growing pump power.

The impact of doping $n_0$ (taken to be equal for both layers at $V_{\rm b}=0$) on the $I-V$ dependence is illustrated in Fig.\ref{fig:currentmap}. The current peaks are shifted towards higher $V_{\rm b}$ values upon  $n_0$ growing more negative [compare to Fig.\ref{fig:current}(c) which shows the $n_0=0$ slice]. The peaks track the $V_{\rm b}$ values at which the emitter layer is bias-doped to  charge neutrality. This is to be expected since the electron-lattice imbalance is maximal at neutrality. The peak-to-valley ratio (PVR) as high as $\sim 5$ can be obtained (see  Fig. \ref{fig:currentmap} inset). For $n_0>0$, in contrast,  the emitter layer is never at charge neutrality for any value of $V_{\rm b}$, resulting in the N-shaped dependence fading out. 

The NDR effect is suppressed under a high bias potential when field emission of carriers with energies below the barrier height overwhelms thermionic emission. As shown in Fig.\ref{fig:current}, the high bias region $eV_{\rm b}/\Delta>1$ is characterized by ${\cal J}$ monotonically growing with increasing $V_{\rm b}$. This behavior arises because lowering the barrier facilitates tunneling and also because growing carrier density results in a faster cooling, thereby reducing the electron-lattice thermal imbalance [see Fig.\ref{fig:current}(c) inset]. 

Next, we proceed to derive a simple criterion for NDR. 
We will focus on the fully-neutral case $n_0=0$ (both graphene layers undoped at $V_{\rm b}=0$) pictured in Fig.\ref{fig:current}. In this case, due to symmetry, we have $T_{1}=T_{2}=T$ and $n_{1}=-n_{2}$ for any $eV_{\rm b}$. 
Further, assuming a small bias and/or a not-too-wide barrier, we can approximate the bias-induced chemical potentials as $\mu_{1,2}\approx \pm eV_{\rm b}/2$. This simple relation is valid for $eV_{\rm b}\ll \kappa(\hbar v)^2/e^2d$, corresponding to the last term in Eq.(\ref{eq:qc}) much smaller than $\mu_1-\mu_2$. Lastly, accounting for the dominant role of thermionic emission, 
we model  transmission as a step function, $|t(\epsilon)|^2 \approx \Gamma^2 \theta (\epsilon - \Delta)$. 
We integrate in Eq.\,(\ref{eq:current}) over energies $\epsilon \ge \Delta \gg \max[k_{\rm B}T,\,V_{\rm b}]$,  approximating $D_{1,2}(\epsilon)$ as a constant and the Fermi distribution tail as $e^{-\beta(\epsilon-\mu)}$. 
This yields Eq.(\ref{eq:thermalcurrent}) to leading order in $V_{\rm b}/\Delta$ and $k_{\rm B}T/\Delta$ with the prefactor $g_0$ given in Eq.(\ref{eq:G0}). 
While the validity of Eq.(\ref{eq:thermalcurrent}) is limited to $d$ which are not too small and also not too large, we find that it predicts NDR in the parameter range close to that found from the full microscopic model used to produce Figs.\ref{fig:current}--\ref{fig:disorder}.

The criterion for NDR can be derived by taking the derivative $d{\cal J}/dV_{\rm b}$ in Eq.\,(\ref{eq:thermalcurrent}) and setting it to zero, giving
\beq
\left( 1+ \frac{1}{\beta \Delta}  \right) \tanh x - \frac{x}{\beta \Delta}=\frac{1}{2\alpha}
,\quad x=\beta e V_{\rm b}/2
, 
\label{eq:condition}
\eeq 
where $\alpha$ is the quantity $-(\Delta/eT)dT/dV_{\rm b}$ introduced above, 
describing the carrier temperature dependence {\it vs.}  $V_{\rm b}$. The $\alpha$ value controls the NDR effect. Maximizing  the left-hand side in $x$ we find the value $f(\lambda)=\lambda^{1/2}-(\lambda-1)\tanh^{-1}\lambda^{-1/2}$ parameterized with $\lambda=1+(\beta\Delta)^{-1}$, which is attained at $x_*=\sinh^{-1}\sqrt{\beta\Delta}$. It is straightforward to check that $f(\lambda)\le 1$ for all $\lambda\ge 1$. This gives the NDR condition $\alpha>1/2$. Below we use this condition, derived for $n_0 = 0$, as an approximation  for the more general case of $n_0 \neq 0$.

To estimate $\alpha$ as a function of the model parameters, it is convenient to factorize $\alpha$ by applying the chain rule as $\alpha=\alpha_{T}\cdot\alpha_{\mu}$, giving
\beq
\alpha_{T}(P)=- \frac{\mu}{T} \frac{dT}{d\mu} 
, \quad \alpha_{\mu}(d) =\frac{\Delta}{e\mu} \frac{d \mu}{d V_{\rm b} } .
\label{eq:alpha2}
\eeq
Here $\alpha_{T}$ depends \textit{only} on the cooling pathways through Eq.\,(\ref{eq:totalpower}), while $\alpha_{\mu}$ depends \textit{only} on the barrier properties through the quantum capacitance effect of Eq.\,(\ref{eq:qc}). Below we use Eq.(\ref{eq:alpha2}) to estimate $\alpha$ and show that the NDR condition $\alpha>1/2$ can be readily met. 

To estimate $\alpha_{T}$ we analyze the degenerate regime $\mu \gg k_{\rm B}T$, where the doping-dependent contributions $P_{\rm ac}$ and $P_{\rm dis}$ dominate over the roughly doping-independent $P_{\rm opt}$ [see  Fig.\ref{fig:disorder}(a) inset]. In this regime, the cooling power behaves as $P_{i}=\gamma_{i}\mu^{a}(T^{b}-T_{0}^{b})$, with $a=4$, $b=1$ for acoustic phonon cooling, and $a=2$, $b=3$ for disorder-assisted cooling (here $\gamma_i$ are constants that depend on the cooling pathways, see Supplement). We assume, for simplicity, that a single cooling pathway dominates over other pathways. Then Eq.\,(\ref{eq:totalpower}) yields
\beq
\alpha_{T}=(a/b)[1-(T_0/T)^b]
\label{eq:alphac}
\eeq
This gives $0<\alpha_{T}< a/b$ with the low and high values corresponding to $T\approx T_{0}$ and $T\gg T_{0}$, respectively. The crossover between these values occurs at a threshold pump power $P_{*}\sim 0.5\,{\rm \mu W/\mu m^2}$ that marks the onset of the hot-carrier regime under typical experimental conditions. We define $P_{*}$ as the value for $P$ at typical carrier densities $n \sim 10^{12}\,{\rm cm}^{-2}$, $T_0=300\,{\rm K}$, and $k_{\rm F} \ell=100$ such that $(T-T_{0})/T_{0} = 0.1$. This yields the above $P_{*}$ value. Maximum $\alpha_{T}$ values found from Eq.(\ref{eq:alphac}) are $0.6$ and $4$ for the $P_{\rm dis}$ and $P_{\rm ac}$ pathways, respectively. 
 
Next, we estimate $\alpha_{\mu}$ as a function of the barrier width. From Eq.\,(\ref{eq:qc}), specializing to the case $n_{0}=0$, we find
\beq
\alpha_{\mu} =\frac{\Delta/2\mu}{1+(4e^2/\kappa \hbar v) k_{\rm F} d} \sim \frac{6.5}{1+0.3 \, d [{\rm nm}]}.
\label{eq:alphad}
\eeq
where $k_{\rm F}=\mu/\hbar v$. Here we have used the hBN barrier value $\Delta\sim 1.3\,{\rm eV}$ and $\mu \sim 0.1\,{\rm eV}$ for typical bias-induced doping. This gives limiting values $\alpha_{\mu}(d\ll d_{*})\approx 6.5 $ and $\alpha_{\mu}(d \gg d_{*})=0$, with the crossover value $d_*\sim 20\,{\rm nm}$. From the above we see that the quantity $\alpha = \alpha_T \cdot \alpha_{\mu}$ can reach values as high as $\alpha\sim 25$.

While the NDR criterion $\alpha>1/2$ is insensitive to the cooling mechanism (so long as it is a strong function of carrier density, as discussed above), the form of the $I-V$ dependence may reflect the cooling mechanism specifics. This is illustrated in Fig.\ref{fig:disorder}(b) for the $P_{\rm dis}$ mechanism. In particular, we consider the bias $V_{\rm pk}$ where the current peaks. From Eq.\,(\ref{eq:condition}) we estimate $eV_{\rm pk}\approx x_* k_{\rm B} T$. When the $P_{\rm dis}$ mechanism
dominates ($k_{\rm F} \ell<10^{3}$) and $T\gg T_{0}$, a power-law relation is obtained: 
$V_{\rm pk}\propto  \left[k_{\rm F} \ell \,  P/\mu^2 \right]^{1/3}$. Similar arguments lead to a disorder-controlled peak-to-valley ratio (PVR). This behavior is illustrated in Fig.\ref{fig:disorder}(b) inset. 

We note that the NDR features may be somewhat smeared out by statistical fluctuations induced by disorder or inhomogeneities. However, we do not expect these effects to destroy NDR. Indeed, optical heating occurs in  ${\rm \mu m}$-wide areas and a typical carrier density is $10^{12}\,{\rm cm}^{-2}$. At the same time, charge inhomogeneity lengthscales in graphene/hBN are a few tens of nm, whereas typical density fluctuations are as low as $\sim 10^{11}\,{\rm cm}^{-2}$ \cite{chargeinhomogeneity,chargeinhomogeneity2}.

Summing up, vertically-stacked graphene heterostructures afford a platform to realize and explore a  range of interesting optoelectronic phenomena due to photogenerated hot carriers. One such phenomenon is the light-induced NDR effect discussed above, manifesting itself through the $I-V$ dependence, acquiring an N-shaped character under optical pumping. Vertical heterostructures use the full graphene area as a photoactive region, and possess a large degree of tunability. These properties make the NDR effect potentially useful for designing new types of optical switches and photodetectors. Our estimates show that the NDR regime, facilitated by  graphene's unique optical and thermal properties, can be readily accessed in wide-barrier heterostructures.

\section{Acknowledgments}

We thank N. Gabor and J. C. W. Song for useful discussions. This work was supported as part of the Center for Excitonics, an Energy Frontier Research Center funded by the U.S. Department of Energy, Office of Science, Basic Energy Sciences under Award No. DE-SC0001088 (LL), and by the National Science Foundation Grant NSF/DMR1004147 (JFRN and MSD).

\section{Supplement: Electronic Cooling Pathways}

Here we summarize the main results on electron-lattice cooling in graphene, following Refs. \cite{coolingmcd,coolingdassarma,disordercooling}. Electron cooling in graphene is usually assumed to be dominated by three main mechanisms: acoustic and optical phonon emission \cite{coolingmcd,coolingdassarma}, and disorder-assisted acoustic phonon emission (``supercollisions'')\cite{disordercooling}. We use the two-temperature model describing the electron and lattice subsystems by two different temperatures, $T$ and $T_0$. This model is valid when the electronic system is thermalized quickly due to fast carrier scattering, whereas the electron-lattice cooling occurs on a longer time scale. 

The contribution to cooling power due to acoustic phonons, obtained for pristine graphene, is given by \cite{coolingmcd}
\beq
P_{\rm ac}=\gamma_{\rm ac}(T-T_{0})\int_{0}^{\infty} d\nu \nu^{3} 4\left[f(\nu)+1-f(-\nu)\right],
\label{eq:Pac}
\eeq
where $f(\epsilon)=[e^{\beta(\epsilon - \mu)}+1]^{-1}$ and the quantity in the prefactor $\gamma_{\rm ac}=\frac{\hbar D^2 k_{\rm B}}{8\pi \rho (\hbar v )^{6}}$ depends on the electron-phonon coupling strength \cite{coolingmcd}. Here $D$ is the deformation potential, $\rho$ is the mass density of the graphene monolayer. 
An explicit dependence on chemical potential $\mu$ can be obtained in the degenerate limit $\beta \mu \gg 1$. In this case, the integral in Eq.\,(\ref{eq:Pac}) yields
\beq
P_{\rm ac} = \gamma_{\rm ac} \mu^{4}(T - T_{0}) .
\eeq
This contribution to cooling, due to its strong dependence on $\mu$, becomes very small near the Dirac point. 

Disorder-assisted acoustic-phonon cooling originates from electron-phonon scattering in the presence of disorder, such that part of phonon momentum is absorbed by disorder. Evaluated for a short-range disorder model, this mechanism yields cooling power \cite{disordercooling} 
\beq
P_{\rm dis}=\gamma_{\rm dis}  \mu^2(T^{3}-T_{0}^{3}) ,\quad \gamma_{\rm dis}=\frac{2 D^2 k_{\rm B}^{3}}{\rho s^2 \hbar (\hbar v )^{4}k_{\rm F} l},
\label{eq:Pdis}
\eeq
where $k_{\rm F} l $ is the dimensionless disorder mean free path parameter, $s$ is the speed of sound, and the degenerate limit $\beta \mu \gg 1$ is assumed. The quadratic dependence of $P_{\rm dis}$ on $\mu$ means that this contribution can win over $P_{\rm ac}$ near the Dirac point. 

Using values of $D\approx 20\,{\rm eV}$, $s\approx 2\cdot 10^4\,{\rm m/s}$, $\rho \approx 7.6 \cdot 10^{-11}\,{\rm kg/cm^2}$, we estimate $\gamma_{\rm ac} \sim 0.5\,{\rm \mu W/ \mu m^2 K(eV)^{4}}$ and $\gamma_{\rm dis} \sim 5 \cdot 10^{-4}\,{\rm \mu W/(eV)^2 K^{3}\mu m^2}$.

Lastly, the cooling power for the optical phonon pathway equals \cite{coolingmcd}
\beq
P_{\rm opt}=\frac{\hbar (\hbar \omega_{0})^{3}}{4 \pi \rho a^{4}(\hbar v )^2}\left[ N_{\rm el}(\omega_{0}) - N_{\rm ph}(\omega_{0})\right] \mathcal{F}(T,\mu),
\eeq
where a flat optical phonon dispersion with $\hbar \omega_{0}=0.2$ eV is assumed, and $a=1.42\,{\rm \AA}$ is the interatomic distance. The quantities  $N_{\rm el}(\omega_{0})$ and $N_{\rm ph}(\omega_{0})$ represent the Bose distribution $[e^{\beta\hbar\omega_0}-1]^{-1}$ evaluated at the electron and lattice temperature,  $T$  and $T_{0}$, respectively.
The quantity $\mathcal{F}(T,\mu)$ is a dimensionless integral 
\beq
\mathcal{F}(T,\mu)=\int_{-\infty}^{\infty} dx |x(x-1)|\left[ f(\hbar \omega_{0} (x-1)) -f(\hbar \omega_{0} x)\right],
\label{eq:foptical}
\eeq
with $f(\epsilon)$ defined above directly after Eq.(\ref{eq:Pac}). 
For weak doping, $\mu\ll\hbar\omega_0$, and $k_{\rm B}T \ll \hbar \omega_{0}$, we can approximate $f(\epsilon)$ by a step function $\theta(-\epsilon)$. Integration in Eq.\,(\ref{eq:foptical}) then yields $\mathcal{F}(T,\mu) \approx 1/6 $, giving 
\beq
P_{\rm opt} = \gamma_{\rm opt} \left[ N_{\rm el}(\omega_{0}) - N_{\rm ph}(\omega_{0})\right] 
,\quad \gamma_{\rm opt}=\frac{\hbar (\hbar \omega_{0})^{3}}{24 \pi \rho a^{4} (\hbar v )^2}.
\eeq
At temperatures $T,\,T_0\ll\hbar\omega_0$ this gives a simple exponential dependence:
\beq
P_{\rm opt} = \gamma_{\rm opt} \left[ e^{-\beta\hbar\omega_{0}} - e^{-\beta_0\hbar\omega_{0}}\right] 
.
\eeq
The  exponential temperature dependence in $P_{\rm opt} $ makes this contribution small at low temperatures. However, the value $P_{\rm opt} $ grows rapidly with tempertaure,  becoming large at temperatures $k_{\rm B}T\gtrsim 0.2\hbar\omega_0$. In addition, even at low temperatures, the effect of $P_{\rm opt} $ can be important near the Dirac point, where other contributions are small since $P_{\rm ac}\propto \mu^4$, $P_{\rm dis}\propto\mu^2$.

Using the above results we estimate the carrier densities that mark the onset of the hot-carrier regime under typical experimental conditions. Reaching these carrier densities by adjusting by the gate potential and voltage bias  triggers the NDR regime (see main text). Below we assume, for concreteness, that electron-lattice cooling is dominated by the disorder-assisted mechanism. Taking a typical pump power value $P\sim 10\,{\rm \mu W/\mu m^2}$, equating it to $P_{\rm dis}$, and using values $k_{\rm F} l \sim 100$ and $T=2 T_{0}\sim 600\,{\rm K}$, Eq.\,(\ref{eq:Pdis}) yields typical carrier densities on the order of $n\sim 10^{12}{\rm cm}^{-2}$. Similar values are obtained for the acoustic phonon mechanism in the absence of disorder. The hot-carrier transport regime is therefore realized when the carrier density is $n \lesssim 10^{12}\,{\rm cm}^{-2}$, whereas  field emission dominates for $n \gtrsim 10^{12}\,{\rm cm}^{-2}$. Such doping values are routinely obtained in graphene systems through electrostatic doping, indicating that the densities required for NDR are easily reachable under realistic experimental conditions.


\begin{thebibliography}{99}
\bibitem{thermionic3D}
Our expression for thermionic current density, Eq.(\ref{eq:thermalcurrent}), features a dependence on the external bias and a power-law temperature dependence distinct from that known for the 3D case, where ${\cal J}_{\rm th}\propto T^2 e^{-\Delta/k_{\rm B}T}$. This reflects the difference in the transport processes at the interface.
Conventional thermionic emission (3D) is described as the flux of free particles with energies $\epsilon>\Delta$ in the direction normal to the surface. In contrast, in our case transport depends on elastic scattering mechanisms at the graphene/hBN interface generating an out-of-plane current, as described in the text.\relax
\end{thebibliography}
\end{document}